\documentclass[aps,pra,twocolumn,showemail,groupedaddress,showpacs]{revtex4-2}

\usepackage[a4paper,margin=0.75in]{geometry}
\usepackage{amsmath,amssymb}
\usepackage{graphicx}
\usepackage{hyperref}
\usepackage{graphics}
\usepackage{float}
\usepackage{placeins}
\usepackage{orcidlink}
\hypersetup{pdfborder = {0 0 0},colorlinks=true,citecolor=magenta,linkcolor=blue,urlcolor=violet}

\usepackage{xcolor}
\usepackage{soul}

\usepackage[normalem]{ulem}
\begin{document}
\title{Quantification of Entanglement in Three-Qubit Systems}
\author{Zikra Aram\,\orcidlink{0009-0006-7381-8913}}
\email{zikraaram02@gmail.com}
\affiliation{Department of Physics, Jamia Millia Islamia, New Delhi, India.}
\affiliation{Department of Physics, Indian Institute of Technology, Patna, India.}\thanks{Present affiliation} 
%\date{}

\begin{abstract}
This study presents an analytical investigation of entanglement
quantification in three-qubit pure states using the Ac\'in canonical
representation. A global concurrence measure is constructed from the
one-qubit-versus-rest bipartite concurrences of the three-qubit
system. A closed-form expression for this measure is derived in terms
of the canonical parameters of a general three-qubit pure state.
Its behavior is then examined for representative product,
biseparable, GHZ-type, and W-type states. The measure is further
compared with the three-tangle and tripartite negativity to examine
their respective sensitivities to different forms of multipartite
entanglement. The results show that the global concurrence remains
sensitive to bipartite entanglement in biseparable states, whereas
tripartite negativity and the three-tangle vanish when genuine
tripartite entanglement is absent. These results demonstrate the
complementary information provided by the different measures in
characterizing entanglement in three-qubit pure states.
\end{abstract}

\maketitle

% ==================================================
\section{Introduction}
% ==================================================

Quantum entanglement is one of the most distinctive and non-classical features of quantum mechanics and plays a fundamental role in quantum information science \cite{Nielsen2010,Horodecki2009}. The conceptual foundations of entanglement can be traced back to the seminal work of Einstein, Podolsky, and Rosen (EPR), who questioned the completeness of quantum mechanics and introduced what is now known as the EPR paradox \cite{EPR1935}. In response to this debate, Bell derived inequalities that enabled experimental tests distinguishing quantum mechanics from local hidden variable theories \cite{Bell1964}. Subsequent experimental violations of Bell inequalities confirmed the nonlocal nature of quantum correlations and established entanglement as a genuine physical phenomenon rather than a philosophical artifact \cite{Aspect1982,Hensen2015}.

Beyond its foundational importance, entanglement has become recognized as a key resource for quantum technologies \cite{Nielsen2010,Bennett2000}. It underpins a variety of quantum information processing tasks, including quantum cryptography \cite{Ekert1991}, quantum teleportation \cite{Bennett1993}, superdense coding \cite{Bennett1992}, and quantum computation \cite{Shor1997}. These applications demonstrate advantages over classical protocols, thereby motivating extensive research into the characterization, detection, and quantification of entanglement \cite{Horodecki2009,Guhne2009}.

A bipartite quantum state is said to be separable if it can be written as a tensor product of states of its individual subsystems; otherwise, it is entangled \cite{Werner1989}. For pure bipartite states, entanglement is completely characterized by the Schmidt decomposition, which provides a canonical representation and a straightforward method for quantification \cite{Schmidt1907,Peres1996}. As a result, bipartite entanglement is relatively well understood, and several entanglement measures such as entanglement of formation and concurrence have been successfully developed in this setting \cite{Bennett1996,Wootters1998}.

In contrast, multipartite entanglement exhibits a significantly richer and more complex structure \cite{Guhne2010}. In multipartite systems, correlations can be distributed among subsystems in inequivalent ways, making classification and quantification substantially more challenging \cite{Dur2000}. For example, in three-qubit systems, entanglement can belong to inequivalent classes such as Greenberger--Horne--Zeilinger (GHZ) states and W states, which cannot be converted into each other under stochastic local operations and classical communication (SLOCC) \cite{GHZ1989,Dur2000}. This inequivalence highlights the fact that no single scalar quantity is sufficient to fully characterize multipartite entanglement \cite{Guhne2010,Eltschka2012}.

Beyond quantum information theory, nonlinear, quantitative, and
computational approaches have been widely employed to investigate
complex problems across different areas of science and engineering.
Entropy-generation methods have been used to analyze fatigue-damage
evolution in solder alloys, while fuzzy-control techniques have been
applied to gyroscope dynamics
\cite{Long2023SAC305Entropy,Zhang2025Type3Gyroscope}. Optimization
and machine-learning approaches have also been employed in
multi-asteroid exploration missions and femtosecond-laser trepan
drilling
\cite{Huo2023ElectricSails,LaserDrilling2022MLGA}. In photonics and
sensing, related quantitative approaches have been applied to
high-$Q$ Fano-resonant structures and multi-FRET-based fluorescence
detection systems
\cite{Peng2025Fano,Xiao2024AFPmFRET}. Complex-network and
data-driven approaches have further been explored for
microRNA--disease association prediction and multimodal sentiment
analysis
\cite{Li2022MiRNANetwork,Jiang2026ThreeWD}. Machine-learning and
multiwavelength data-analysis techniques have also been developed for
solar-flare prediction and gamma-ray blazar identification
\cite{Yang2025SolarFlares,Xiang2026Blazars}. Related dynamical and
probabilistic approaches have been investigated in information
propagation and stabilizer testing
\cite{Tian2026InformationPropagation,Levy2025StabilizerCLT}.
Quantitative and optimization-based methods have additionally been
applied to structural reliability and spacecraft orbit maintenance
problems
\cite{QuantumMapping2025Reliability,Ding2026OrbitMaintenance}.
In condensed-matter and photonic systems, related studies have
investigated quantum anomalous Hall effects, quantum-dot light-emitting
devices, and non-Hermitian photonic phenomena
\cite{Qi2016QAHE,Li2025PureRedPeLED,Zhang2025NonHermitianDrift}.
Experimental quantum-optical approaches have also been used to
visualize quantum erasure with hybrid entanglement
\cite{Yu2026QuantumErasure}.

\ Several approaches have been proposed to quantify multipartite entanglement, including concurrence-based measures, three-tangle, and other polynomial invariants \cite{Wootters1998,Coffman2000}. In particular, Coffman, Kundu, and Wootters introduced the concept of monogamy of entanglement and the three-tangle as a measure of genuine tripartite entanglement \cite{Coffman2000}. Nevertheless, a complete and universally accepted framework for multipartite entanglement quantification continues to be a subject of study \cite{Guhne2010,Amico2008,Dong2024}.
%quez2024Enrí, Vathsan2025

\ One useful strategy for investigating multipartite entanglement is to analyze entanglement across different bipartitions of the system. \cite{Plenio2007}. By examining the reduced density matrices associated with various subsystem partitions, one can extract information about how entanglement is distributed among the parties \cite{Horodecki2009,Rungta2001}. {Bhaskara and Panigrahi \cite{Bhaskara2017} proposed a particularly convenient method by defining a \emph{global concurrence} which is just the sum of the concurrences of all the bipartitions.} They evaluated the global concurrence for a general 3-qubit entangled state. The number of parameters, however, can be reduced by writing the 3-qubit entangled state as a generalized Schmidt decomposition (GSD) form. A convenient
parametrization of the GSD state is given by the Ac\'in canonical form \cite{Acin2000}. This decomposition removes redundancies arising from local unitary transformations and provides a convenient starting point for the analysis of tripartite entanglement. The five parameters prove to be useful in classifying 3-qubit entangled states \cite{Sabin2008}.

{Driven by these considerations, this study concentrates on quantifying entanglement in three-qubit pure states, utilizing the global concurrence introduced by Bhaskara and Panigrahi, while applying the GSD representation of the state. %This procedure enabled us to carry out a complete SLOCC-class analysis of the Bhaskara-Panigrahi global concurrence in the Acín 
%canonical form of the 3-qubit state.} 
This procedure enabled us to evaluate the measure on representative Ac\'in parameter families associated with standard three-qubit entanglement types.}
Our results contribute to the broader effort to better understand the structure and quantification of multipartite entanglement in quantum information theory. {A particular strength of the analysis is obtaining closed-form expressions for the entanglement measure for the general 3-qubit entangled state and for various classes.}

\section{Schmidt decomposition and entanglement}

The Schmidt decomposition \cite{Schmidt1907} provides a canonical representation for any bipartite pure state. For a pure state $|\psi_{AB}\rangle$ in the Hilbert space $\mathcal{H}_A \otimes \mathcal{H}_B$, there exist orthonormal bases $\{|i_A\rangle\}$ and $\{|i_B\rangle\}$ such that
\begin{equation}
|\psi_{AB}\rangle = \sum_i \lambda_i |i_A\rangle \otimes |i_B\rangle,
\label{eq:schmidt}
\end{equation}
where $\lambda_i \ge 0$ are the Schmidt coefficients satisfying $\sum_i \lambda_i^2 = 1$. The number of non-zero Schmidt coefficients is called the Schmidt rank.

A bipartite pure state is separable if and only if its Schmidt rank equals one, i.e., $|\psi_{AB}\rangle = |\phi_A\rangle \otimes |\chi_B\rangle$. For entangled states, the Schmidt rank is greater than one. Several entanglement measures for bipartite pure states exist. The concurrence for two-qubit systems \cite{Wootters1998} provides a particularly useful quantification. For a general two-qubit pure state
\begin{equation}
|\psi\rangle = a|00\rangle + b|01\rangle + c|10\rangle + d|11\rangle,
\end{equation}
the concurrence is given by
\begin{equation}
C(|\psi\rangle) = 2|ad - bc|,
\label{eq:concurrence}
\end{equation}
which vanishes for separable states ($ad = bc$) and reaches unity for maximally entangled states.
For bipartite pure states, the coherence \cite{Baumgratz2014} of the state in the Schmidt basis turns out to be a good measure of entanglement \cite{Pathania2022}.

While bipartite entanglement is well understood, multipartite entanglement exhibits significantly richer structure \cite{Dur2000,Sabin2008}. The Schmidt decomposition does not generalize straightforwardly to systems with three or more parties \cite{Pati2000,Carteret2000}. In multipartite systems, entanglement can be distributed among subsystems in inequivalent ways, and different types of entanglement can coexist.

For three-qubit systems, states can be classified into inequivalent entanglement classes under stochastic local operations and classical communication (SLOCC) \cite{Dur2000}. The two most prominent classes are:
\begin{itemize}
    \item \textbf{GHZ class:} Exemplified by $|GHZ\rangle = \frac{1}{\sqrt{2}}(|000\rangle + |111\rangle)$, representing genuine tripartite entanglement.
    \item \textbf{W class:} Exemplified by $|W\rangle = \frac{1}{\sqrt{3}}(|001\rangle + |010\rangle + |100\rangle)$, with different entanglement distribution properties.
\end{itemize}
These classes cannot be interconverted by local unitary operations, highlighting the complexity of multipartite entanglement.

To facilitate systematic analysis of three-qubit pure states, Ac\'in et al.\ \cite{Acin2000} introduced a canonical form obtained through local unitary transformations. Any three-qubit pure state can be written as
\begin{equation}
|\psi\rangle = \lambda_0 |000\rangle + \lambda_1 e^{i\phi}|100\rangle + \lambda_2 |101\rangle + \lambda_3 |110\rangle + \lambda_4 |111\rangle,
\label{eq:acin_form}
\end{equation}
where the parameters satisfy:
\begin{equation}
\lambda_i \ge 0, \quad \sum_{i=0}^4 \lambda_i^2 = 1, \quad 0 \le \phi \le \pi.
\label{eq:acin_conditions}
\end{equation}
This representation reduces the parameter space from the general $2^3 = 8$ complex amplitudes (with normalization) to five real non-negative amplitudes and one phase. The canonical form removes redundancies arising from local unitary transformations and provides a convenient parametrization for analyzing entanglement properties.

In the Ac\'in canonical form, one can systematically examine the reduced density matrices corresponding to different bipartitions of the three-qubit system. By analyzing these reduced states across various bipartitions, one can study how entanglement is distributed within the tripartite system. This bipartition-based approach, extending concurrence-related ideas from bipartite to multipartite systems, forms the foundation of the analysis presented in the following section.

\section{entanglement measure for three-qubit systems}

\subsection{Motivation and approach}

As discussed in the previous section, the Schmidt decomposition provides a complete characterization of bipartite entanglement, but does not generalize straightforwardly to multipartite systems. To address this limitation, we adopt a bipartition-based approach. In this method, we analyze the reduced density matrix of each subsystem separately. The mixedness of each subsystem arises due to its entanglement with the remaining subsystems.

We study the global concurrence measure of entanglement as a sum of the \emph{generalized concurrence} of all the bipartitions \cite{Bhaskara2017}. The generalized concurrence of each bipartition can be written in terms of the determinant of the reduced density matrices of an individual qubit, after tracing over the other two qubits. This has been done for three qubits earlier \cite{Bhaskara2017}, but it involves eight parameters, and  characterizing the three-qubit entanglement becomes a difficult task.  In the following we derive the analytical expressions of this measure using the Ac\'in canonical form of the Schmidt decomposition of the three-qubit state. The advantage here that with the reduced number of parameters, it becomes possible to characterize the entanglement of various classes of the three-qubit state.

\subsection{Reduced density matrices and determinants}

We start by calculating the reduced density matrices for different subsystems using the Ac\'in canonical form. 
Consider the Ac\'in canonical form of a general three-qubit pure state:
\begin{equation}
|\psi\rangle = \lambda_0|000\rangle + \lambda_1 e^{i\phi}|100\rangle + \lambda_2|101\rangle + \lambda_3|110\rangle + \lambda_4|111\rangle,
\label{eq:acin_state}
\end{equation}
with the normalization condition
\begin{equation}
\sum_{i=0}^{4} \lambda_i^2 = 1,
\label{eq:normalization}
\end{equation}
where $\lambda_i \geq 0$ and $\phi \in [0, \pi]$.

\subsubsection{Reduced density matrix of subsystem A}

{The reduced density matrix of subsystem $A$ is obtained by tracing over subsystems $B$ and $C$:}
\begin{equation}
\rho_A = \operatorname{Tr}_{BC}(|\psi\rangle\langle\psi|).
\end{equation}

{Therefore,}
\begin{equation}
\rho_A = 
\begin{pmatrix}
\lambda_0^2 & \lambda_0\lambda_1 e^{-i\phi} \\
\lambda_0\lambda_1 e^{i\phi} & \lambda_1^2 + \lambda_2^2 + \lambda_3^2 + \lambda_4^2
\end{pmatrix}.
\label{eq:rho_A}
\end{equation}

{Using the normalization condition, the diagonal element can be written as 
$\lambda_1^2 + \lambda_2^2 + \lambda_3^2 + \lambda_4^2 = 1 - \lambda_0^2$.}

{The determinant is}
\begin{equation}
\det\rho_A = \lambda_0^2(\lambda_1^2 + \lambda_2^2 + \lambda_3^2 + \lambda_4^2) 
- \lambda_0^2\lambda_1^2 = \lambda_0^2(\lambda_2^2 + \lambda_3^2 + \lambda_4^2).
\label{eq:det_rho_A}
\end{equation}

{Equivalently,}
\begin{equation}
\det\rho_A = \lambda_0^2(1 - \lambda_0^2 - \lambda_1^2).
\label{eq:det_rho_A_alt}
\end{equation}

\subsubsection{Reduced density matrix of subsystem B}

{Tracing over subsystems $A$ and $C$, we obtain}
\begin{equation}
\rho_B = \operatorname{Tr}_{AC}(|\psi\rangle\langle\psi|).
\end{equation}

{Therefore,}
\begin{equation}
\rho_B = 
\begin{pmatrix}
\lambda_0^2 + \lambda_1^2 + \lambda_2^2 & \lambda_1\lambda_3 e^{i\phi} + \lambda_2\lambda_4 \\
\lambda_1\lambda_3 e^{-i\phi} + \lambda_2\lambda_4 & \lambda_3^2 + \lambda_4^2
\end{pmatrix}.
\label{eq:rho_B}
\end{equation}

{The determinant is computed as}
\begin{equation}
\det\rho_B = (\lambda_0^2 + \lambda_1^2 + \lambda_2^2)(\lambda_3^2 + \lambda_4^2) 
- |\lambda_1\lambda_3 e^{i\phi} + \lambda_2\lambda_4|^2.
\end{equation}

{After simplification, we obtain}
\begin{equation}
\det\rho_B = \lambda_0^2\lambda_3^2 + \lambda_0^2\lambda_4^2 + \lambda_1^2\lambda_4^2 
+ \lambda_2^2\lambda_3^2 - 2\lambda_1\lambda_2\lambda_3\lambda_4\cos\phi.
\label{eq:det_rho_B}
\end{equation}

\subsubsection{Reduced density matrix of subsystem C}

{Similarly, tracing over subsystems $A$ and $B$,}
\begin{equation}
\rho_C = \operatorname{Tr}_{AB}(|\psi\rangle\langle\psi|).
\end{equation}

{Thus,}
\begin{equation}
\rho_C = 
\begin{pmatrix}
\lambda_0^2 + \lambda_1^2 + \lambda_3^2 & \lambda_1\lambda_2 e^{i\phi} + \lambda_3\lambda_4 \\
\lambda_1\lambda_2 e^{-i\phi} + \lambda_3\lambda_4 & \lambda_2^2 + \lambda_4^2
\end{pmatrix}.
\label{eq:rho_C}
\end{equation}

{The determinant is}
\begin{equation}
\det\rho_C = (\lambda_0^2 + \lambda_1^2 + \lambda_3^2)(\lambda_2^2 + \lambda_4^2) 
- |\lambda_1\lambda_2 e^{i\phi} + \lambda_3\lambda_4|^2.
\end{equation}

{After expanding and simplifying,}
\begin{equation}
\det\rho_C = \lambda_0^2\lambda_2^2 + \lambda_0^2\lambda_4^2 + \lambda_1^2\lambda_4^2 
+ \lambda_2^2\lambda_3^2 - 2\lambda_1\lambda_2\lambda_3\lambda_4\cos\phi.
\label{eq:det_rho_C}
\end{equation}

\subsection{Analytical expression for global concurrence}

To quantify the total entanglement contribution from all bipartitions of the system, we adopt the global concurrence measure proposed by Bhaskara and Panigrahi \cite{Bhaskara2017} which, {in its normalized form,} is defined as

\begin{equation}
E(|\psi\rangle) = \frac{2}{3}\left( \sqrt{\det\rho_A} + \sqrt{\det\rho_B} 
+ \sqrt{\det\rho_C} \right).
\label{eq:E_definition}
\end{equation}

where $\rho_A$, $\rho_B$, and $\rho_C$ are the reduced density matrices of subsystems A, B, and C, respectively. This expression captures the total entanglement contribution arising from all bipartitions of the system. Each term represents the contribution from one bipartition. {Thus, $0\leq E\leq1$.}

{Substituting the determinants from Eqs.~\eqref{eq:det_rho_A_alt}, 
\eqref{eq:det_rho_B}, and \eqref{eq:det_rho_C}, we obtain the explicit expression}
\begin{equation}
\begin{split}
E = \frac{2}{3} \Bigg[ & \lambda_0\sqrt{(1 - \lambda_0^2 - \lambda_1^2)} \\
& + \sqrt{\lambda_0^2\lambda_3^2 + \lambda_0^2\lambda_4^2 + \lambda_1^2\lambda_4^2 
+ \lambda_2^2\lambda_3^2 - 2\lambda_1\lambda_2\lambda_3\lambda_4\cos\phi} \\
& + \sqrt{\lambda_0^2\lambda_2^2 + \lambda_0^2\lambda_4^2 + \lambda_1^2\lambda_4^2 
+ \lambda_2^2\lambda_3^2 - 2\lambda_1\lambda_2\lambda_3\lambda_4\cos\phi} \Bigg].
\end{split}
\label{eq:E_explicit}
\end{equation}
with the constraints
\begin{equation}
\lambda_i \geq 0, \quad \sum_{i=0}^{4} \lambda_i^2 = 1, \quad \phi \in [0, \pi].
\label{eq:constraints}
\end{equation}

The measure $E$ is expressed explicitly in terms of the five amplitude parameters and the phase parameter of the Ac\'in canonical form. In the following section, we investigate the properties of this measure and evaluate it for various classes of three-qubit entangled states.

\subsection{General properties}

Before presenting numerical results, we first establish several important properties of the measure $E$ defined in Eq.~(\ref{eq:E_explicit}).

\paragraph{Separability.} For a fully separable state $|\psi\rangle = |\psi_A\rangle \otimes |\psi_B\rangle \otimes |\psi_C\rangle$, all reduced density matrices are pure, and hence their determinants vanish. This implies $\det(\rho_A) = \det(\rho_B) = \det(\rho_C) = 0$, and consequently $E= 0$.

\paragraph{Non-negativity.} Since $E$ is defined as a sum of square roots of determinants, and all density matrices are positive semidefinite, the measure $E \geq 0$ for all quantum states.

\paragraph{Normalization.} The measure is bounded, with $0 \leq E \leq1$, where the maximum value depends on the entanglement class and structure of the state.

\subsubsection{Relation to bipartite concurrence and LOCC monotonicity}

{The global concurrence can be rewritten in terms of the one-qubit-versus-rest 
bipartite concurrences. For a pure bipartite state with a single qubit on one side, 
the concurrence is given by $C = 2\sqrt{\det\rho}$, where $\rho$ is the reduced density 
matrix of the single qubit. Therefore,}
\begin{equation}
\begin{aligned}
C_{A|BC} &= 2\sqrt{\det\rho_A}, \\
C_{B|AC} &= 2\sqrt{\det\rho_B}, \\
C_{C|AB} &= 2\sqrt{\det\rho_C}.
\end{aligned}
\label{eq:bipartite_concurrences}
\end{equation}

{Substituting into Eq.~\eqref{eq:E_definition}, we obtain}
\begin{equation}
E(|\psi\rangle) = \frac{1}{3}\left( C_{A|BC} + C_{B|AC} + C_{C|AB} \right).
\label{eq:E_as_average}
\end{equation}

{This shows that $E$ is the arithmetic mean of the three one-qubit-versus-rest 
bipartite concurrences. Each of these concurrences is an entanglement monotone under 
LOCC for pure bipartite states~\cite{Vidal2000,Wootters1998}. Since any tripartite LOCC 
protocol is also a valid LOCC protocol for each bipartition, the monotonicity property 
is inherited. For an LOCC transformation}
\begin{equation}
|\psi\rangle \longrightarrow \{p_k, |\psi_k\rangle\},
\end{equation}
{we have}
\begin{equation}
C_{i|\bar{i}}(|\psi\rangle) \geq \sum_k p_k C_{i|\bar{i}}(|\psi_k\rangle), 
\quad i = A, B, C.
\end{equation}

{Averaging over the three bipartitions, we obtain}
\begin{equation}
E(|\psi\rangle) \geq \sum_k p_k E(|\psi_k\rangle).
\label{eq:LOCC_monotonicity}
\end{equation}

{Therefore, $E$ is non-increasing on average under LOCC for pure three-qubit states.}

\textbf{{Important distinction:}} {This LOCC monotonicity should not be 
confused with SLOCC invariance. The quantity $E$ is generally not invariant under 
invertible local filtering operations. Therefore, $E$ is not a SLOCC invariant and 
cannot serve as a complete classifier of SLOCC equivalence classes. Rather, it should 
be interpreted as a local-unitary-invariant and LOCC-monotonic quantifier of the 
average one-qubit-versus-rest entanglement.}

\subsubsection{Local unitary invariance}

{The measure $E$ is invariant under local unitary transformations. If}
\begin{equation}
|\psi'\rangle = (U_A \otimes U_B \otimes U_C)|\psi\rangle,
\end{equation}
{where $U_A$, $U_B$, and $U_C$ are unitary operators, then the reduced density 
matrices transform as}
\begin{equation}
\rho_i' = U_i \rho_i U_i^\dagger, \quad i = A, B, C.
\end{equation}

{Since the determinant is invariant under unitary conjugation,}
\begin{equation}
\det\rho_i' = \det\rho_i,
\end{equation}
{we have}
\begin{equation}
E(|\psi'\rangle) = E(|\psi\rangle).
\label{eq:LU_invariance}
\end{equation}

{Thus, $E$ is a local-unitary invariant.}

\section{Analytical behavior and numerical plots}

To understand the behavior of the global concurrence measure and validate its physical significance, we now examine its analytical properties and numerical characteristics. We analyze the variation of $E$ and its individual components across the parameter space defined by the Ac\'in canonical form.

\subsection{Parametric variation}

\begin{figure}[htbp]
\centering
\includegraphics[width=\columnwidth]{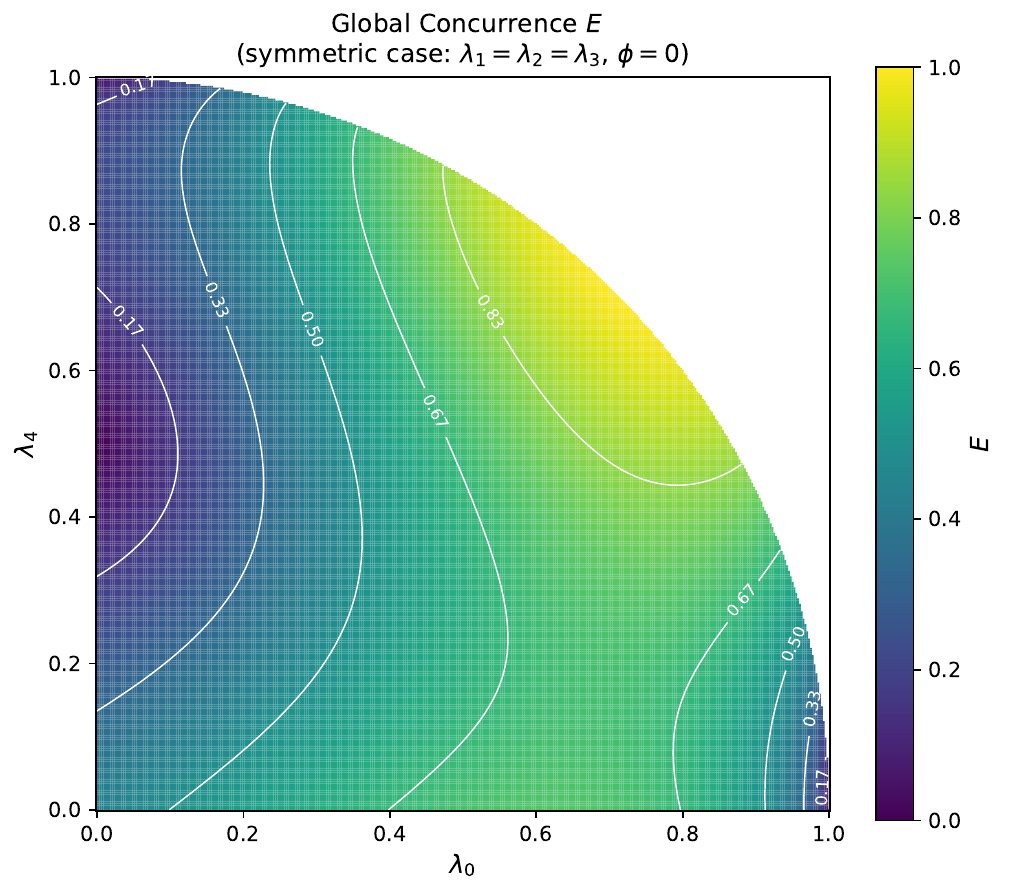}
\caption{Contour plot showing the variation of the global concurrence
$E$ for the symmetric parameter slice
$\lambda_1=\lambda_2=\lambda_3$ and $\phi=0$. The physically allowed
region satisfies $\lambda_0^2+\lambda_4^2\leq1$. The point
$\lambda_0=\lambda_4=1/\sqrt{2}$ corresponds to the GHZ state and
gives the maximum value $E=1$. The endpoints
$(\lambda_0,\lambda_4)=(1,0)$ and $(0,1)$ correspond to the product
states $|000\rangle$ and $|111\rangle$, respectively.}
\label{fig:contour}
\end{figure}

To visualize the behavior of the measure across the parameter space, we plot the variation of $E$ and its individual components for the symmetric case $\lambda_1 = \lambda_2 = \lambda_3$ with the phase fixed at $\phi = 0$ for simplicity in Fig. \ref{fig:contour}.
From the plot, it can be seen that the value of the measure changes smoothly over the allowed region defined by the normalization constraint. The value becomes zero at the product state, where there is no entanglement present, as expected. This behavior indicates that the measure correctly identifies separable states.
The measure takes higher values near the GHZ region, indicating the presence of genuine tripartite entanglement in that state. For the W state, the value is non-zero but smaller compared to the GHZ case, reflecting the different entanglement structure of these inequivalent classes.

\subsection{Discussion}

The analysis presented in this section demonstrates several key features of the global concurrence:

\begin{itemize}
    \item The measure correctly vanishes for separable states.
    \item It takes non-zero values for entangled states, with the magnitude reflecting the degree and type of entanglement.
    
    \item The individual contributions from each bipartition provide insight into how entanglement is distributed across the system.
    \item The measure varies smoothly across the parameter space.
\end{itemize}

These properties indicate that this measure provides a meaningful quantification of entanglement in three-qubit pure states and captures important physical features of multipartite quantum correlations.

\section{Assessment of entanglement for different classes of three-qubit states}

Having derived the general expression for $E$ in Eq.~(\ref{eq:E_explicit}), We now evaluate  \(E\) for representative subclasses of pure three-qubit states written in the Ac\'{\i}n canonical form \cite{Acin2000}. 

\subsection{Classification framework based on Acín et al.}

Following the work of Acín et al.\ \cite{Acin2000}, three-qubit pure states can be systematically classified according to polynomial invariants. These invariants, constructed from the Acín canonical parameters, determine the entanglement structure of the state.

A fundamental invariant in this classification is $J_1$, defined as
\begin{equation}
J_1 = |\lambda_1\lambda_4 e^{i\phi} - \lambda_2\lambda_3|^2.
\label{eq:J1_invariant}
\end{equation}

Additional invariants $J_2, J_3, J_4, J_5$ involve higher-order combinations of the parameters. The vanishing or non-vanishing of these invariants, together with constraints on the Acín parameters $\{\lambda_0, \lambda_1, \lambda_2, \lambda_3, \lambda_4, \phi\}$, uniquely determines the entanglement class \cite{Acin2000,Sabin2008}.

The classification identifies several distinct types:

\begin{table}[t]
\centering
\caption{Representative subclasses in the Ac\'{\i}n parametrization and
the corresponding value of the measure \(E\).}
\label{tab:classsummary}
\renewcommand{\arraystretch}{1.2}
\setlength{\tabcolsep}{4pt}
\begin{tabular}{|c|c|c|}
\hline
\textbf{Subclass} & \textbf{Parameter restriction} & \textbf{Measure \(E\)} \\
\hline
Product state
& \(\lambda_2=\lambda_3=\lambda_4=0\)
& \(0\) \\
\hline
Biseparable state
& \(\lambda_3=\lambda_4=0\)
& \(\dfrac{4}{3}\lambda_0\lambda_2\) \\
\hline
GHZ-type state
& \(\lambda_1=\lambda_2=\lambda_3=0\)
& \(2\lambda_0\lambda_4\) \\
\hline
Standard GHZ state
& \(\lambda_0=\lambda_4=1/\sqrt{2}\)
& \(1\) \\
\hline
W-type state
& \(\lambda_1=\lambda_4=0\)
& Eq.~(\ref{eq:EWfinal}) \\
\hline
Standard W state
& \(\lambda_0=\lambda_2=\lambda_3=1/\sqrt{3}\)
& \(\dfrac{2\sqrt{2}}{3}\) \\
\hline
\end{tabular}
\end{table}

Table~\ref{tab:classsummary} summarizes the behavior of the E for the representative subclasses considered in this work. We now discuss each case in more detail.

We now analyze the behavior of $E$ for each of these classes.

\subsection{Type 1: Product states}

For fully separable product states, the Acín parameters satisfy
\begin{equation}
\lambda_2 = 0, \quad \lambda_3 = 0, \quad \lambda_4 = 0.
\end{equation}
According to the Acín classification \cite{Acin2000}, this corresponds to all polynomial invariants vanishing: $J_i = 0$ for all $i$. The state factorizes as
\begin{equation}
|\psi\rangle = (\lambda_0|0\rangle + \lambda_1 e^{i\phi}|1\rangle)_A \otimes |0\rangle_B \otimes |0\rangle_C,
\end{equation}
where $\lambda_0^2 + \lambda_1^2 = 1$.
Substituting into Eq.~(\ref{eq:E_explicit}), all three terms vanish:
\begin{equation}
 {E_{\mathrm{prod}}}   = 0.
\end{equation}
This confirms that the measure correctly identifies fully separable states with no entanglement.

\subsection{Type 2(a): Biseparable states}

In the Ac\'{\i}n classification~\cite{Acin2000}, Type 2(a) corresponds to
biseparable states for which one of the invariants $J_1$, $J_2$, or
$J_3$ may be nonzero, depending on which qubit is separable, while
$J_4=J_5=0$. For the $B$--$(AC)$ representative considered here,
the conditions $J_2=J_3=J_4=J_5=0$ are replaced by the corresponding
choice of separable qubit, yielding $\lambda_3=\lambda_4=0$. In this
case,
\[
J_1=J_3=J_4=J_5=0,\qquad
J_2=\lambda_0^2\lambda_2^2\neq0,
\]
for a nontrivially entangled two-qubit subsystem.

Biseparable states have the form $|\psi\rangle = |\phi\rangle_B\otimes |\chi\rangle_{AC}$, where two qubits are entangled while the third remains separable. 

 The Acín form reduces to
\begin{equation}
|\psi\rangle = \lambda_0|000\rangle + \lambda_1 e^{i\phi}|100\rangle + \lambda_2|101\rangle,
\end{equation}
with normalization $\lambda_0^2 + \lambda_1^2 + \lambda_2^2 = 1$.

Since qubit B remains in state $|0\rangle$, the state factorizes as
\begin{equation}
|\psi\rangle = |0\rangle_B \otimes (\lambda_0|00\rangle + \lambda_1 e^{i\phi}|10\rangle + \lambda_2|11\rangle)_{AC}.
\end{equation}

\begin{align}
\det\rho_A &= \lambda_0^2\lambda_2^2,\\
\det\rho_B &= 0,\\
\det\rho_C &= \lambda_0^2\lambda_2^2.
\end{align}
Hence,
\begin{equation}
E_{\mathrm{bisep}}
=
\frac{2}{3}\left(\lambda_0\lambda_2+0+\lambda_0\lambda_2\right)
=
\frac{4}{3}\lambda_0\lambda_2.
\label{eq:Ebisepfinal}
\end{equation}

Key observations:
\begin{itemize}
\item $E > 0$ due to bipartite AC entanglement;
\item The result is independent of the phase $\phi$;
\item Although the three-tangle vanishes, $E \neq 0$, indicating that the measure captures bipartite correlations.
\end{itemize}

\subsection{Type 2(b): GHZ-type states}

The GHZ class in the Acín classification \cite{Acin2000} is characterized by $J_4 \neq 0$ with all other invariants vanishing. This class represents genuine tripartite entanglement. The constraint $J_1 = 0$ yields
\begin{equation}
\lambda_1\lambda_4 e^{i\phi} = \lambda_2\lambda_3,
\end{equation}
while the conditions $J_2 = \lambda_0^2\lambda_2^2 = 0$ and $J_3 = \lambda_0^2\lambda_3^2 = 0$ lead to $\lambda_2 = \lambda_3 = 0$ (since $\lambda_0 \neq 0$ for non-trivial states).
The Acín canonical state reduces to
\begin{equation}
|\psi\rangle = \lambda_0|000\rangle + \lambda_4|111\rangle,
\end{equation}
with normalization $\lambda_0^2 + \lambda_4^2 = 1$.
Substituting into Eq.~(\ref{eq:E_explicit}):
\begin{eqnarray}
{E_{\mathrm{GHZ}}} &=& {\frac{2}{3}}
\left[\lambda_0\sqrt{1-\lambda_0^2}
+\lambda_0\lambda_4 +\lambda_0\lambda_4 \right]\nonumber\\
&=& \left[2\lambda_0\lambda_4\right].
\end{eqnarray}

For the maximally entangled GHZ state,
\begin{equation}
|GHZ\rangle = \frac{1}{\sqrt{2}}(|000\rangle + |111\rangle),
\end{equation}
we have $\lambda_0 = \lambda_4 = 1/\sqrt{2}$, yielding
${E_{\mathrm{GHZ}}}= 1$.
This represents the maximum value of the measure, capturing genuine tripartite entanglement.

\subsection{Type 3(a): W-type states}

For a W-type representative in the Ac\'{\i}n form, we take
\begin{equation}
\lambda_1=\lambda_4=0,
\end{equation}
so that
\begin{equation}
|\psi_{W}\rangle
=
\lambda_0|000\rangle+\lambda_2|101\rangle+\lambda_3|110\rangle,
\qquad
\lambda_0^2+\lambda_2^2+\lambda_3^2=1.
\end{equation}
In this case,
\begin{align}
\det\rho_A &= \lambda_0^2({1-\lambda_0^2}),\\
\det\rho_B &= \lambda_0^2\lambda_3^2+\lambda_2^2\lambda_3^2,\\
\det\rho_C &= \lambda_0^2\lambda_2^2+\lambda_2^2\lambda_3^2.
\end{align}
Thus,
\begin{equation}
E_W
=
\frac{2}{3}\left[
\lambda_0\sqrt{1-\lambda_0^2}
+\lambda_3\sqrt{\lambda_0^2+\lambda_2^2}
+\lambda_2\sqrt{\lambda_0^2+\lambda_3^2}
\right].
\label{eq:EWfinal}
\end{equation}

For the standard W state
\begin{equation}
|W\rangle=\frac{1}{\sqrt{3}}\left(|100\rangle+|010\rangle+|001\rangle\right),
\end{equation}
which is locally equivalent to the Ac\'{\i}n representative
\begin{equation}
|W'\rangle=\frac{1}{\sqrt{3}}\left(|000\rangle+|101\rangle+|110\rangle\right),
\end{equation}
we identify
\begin{equation}
\lambda_0=\lambda_2=\lambda_3=\frac{1}{\sqrt{3}},
\qquad
\lambda_1=\lambda_4=0.
\end{equation}
Substitution into Eq.~\eqref{eq:EWfinal} yields
\begin{equation}
E_W=\frac{2\sqrt{2}}{3}\approx 0.943.
\end{equation}

Therefore, the measure remains nonzero for W-type states, again reflecting the presence of distributed bipartite entanglement across the three one-qubit cuts.

% Optional figure for W subclass
% \begin{figure}[t]
% \centering
% \includegraphics[width=0.58\textwidth]{w_plot.pdf}
% \caption{Behavior of \(E_W\) for the W-type subclass.}
% \label{fig:w}
% \end{figure}

Important observations:
\begin{itemize}
\item The expression is independent of the phase $\phi$;

\item $E(W) > 0$ despite vanishing three-tangle, indicating sensitivity to bipartite-type correlations that characterize the W class.
\end{itemize}

\subsection{Summary of the class-wise behavior}

From the above results, the following observations can be made:
\begin{itemize}
    \item \(E=0\) for fully separable product states.
    \item \(E\neq 0\) in general for biseparable states, showing that the quantity is sensitive to bipartite entanglement contributions.
    \item For GHZ-type states, \(E\) reaches the maximal value \(E=1\) for the standard GHZ state.
    \item For the standard W state, \(E=2\sqrt{2}/3\), so the measure also detects nontrivial entanglement in the W-type subclass.
\end{itemize}

\subsection{Numerical plots}

\begin{figure}[htbp]
\centering
\includegraphics[width=\columnwidth]{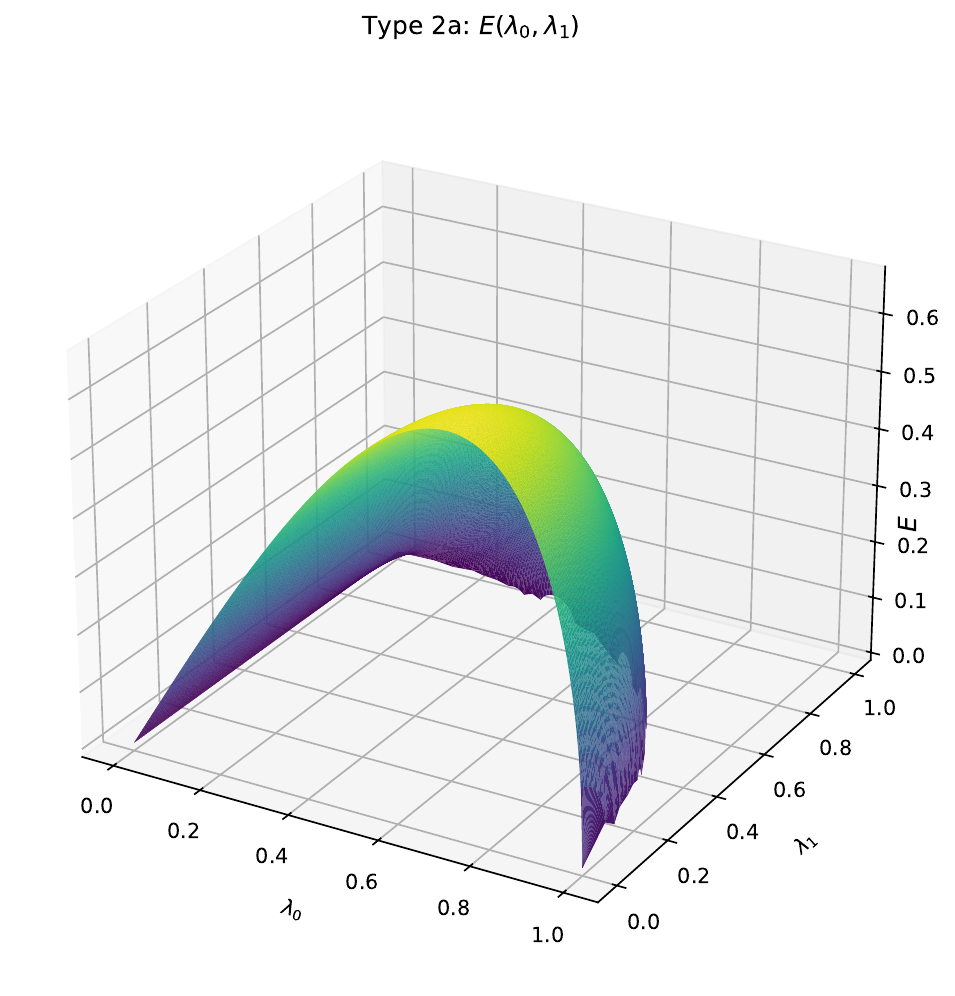}
\caption{Variation of the global concurrence $E$ for Type 2(a)
biseparable states as a function of the canonical parameters
$\lambda_0$ and $\lambda_1$. The measure is nonzero in the interior
of the parameter region and vanishes at the boundaries corresponding
to product states.}
\label{fig:type2a}
\end{figure}

To visualize the behavior of the measure across the Acín parameter space, we present contour plots for representative classes.
Fig.~\ref{fig:type2a} shows the variation of $E$ for Type 2(a) biseparable states as a function of different parameter combinations. The measure vanishes at boundary regions where coefficients approach zero, consistent with the limit toward product states.

\begin{figure}[htbp]
\centering
\includegraphics[width=\columnwidth]{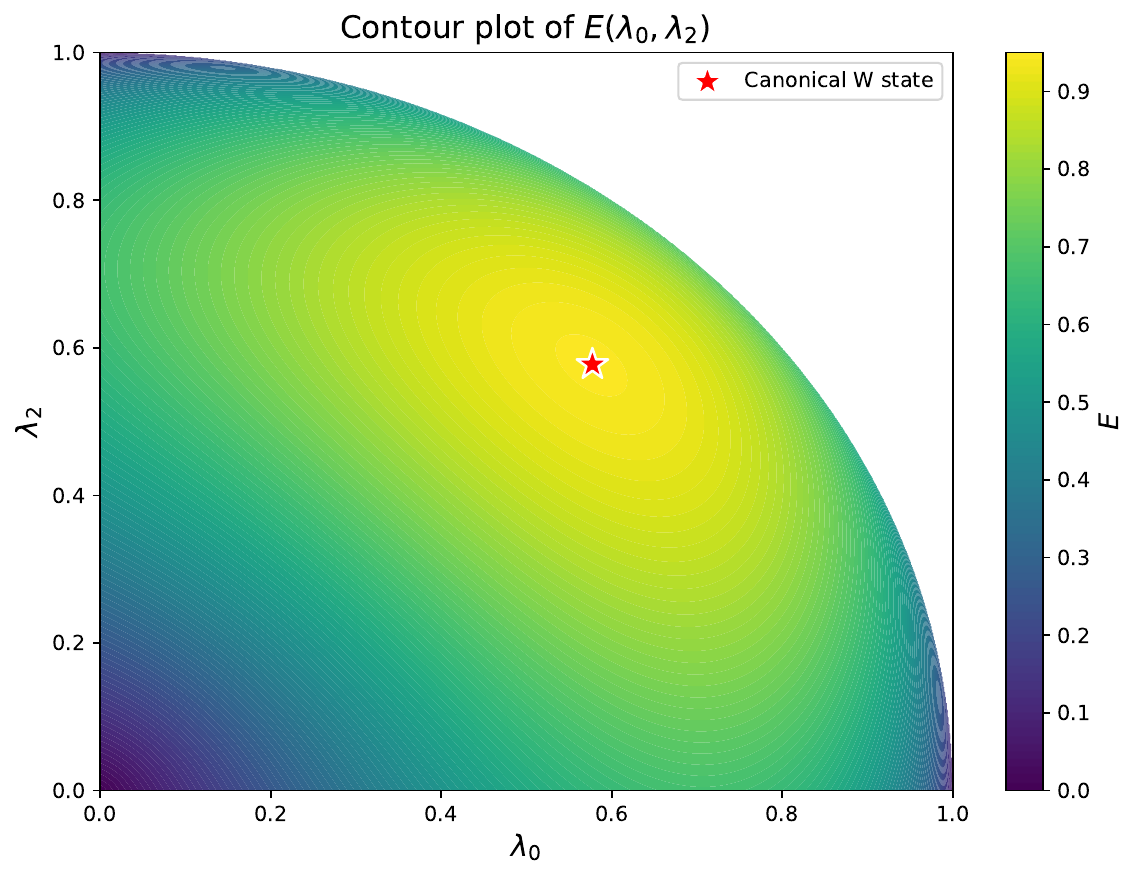}
\caption{Variation of the global concurrence $E$ for Type 3(a)
W-type states as a function of $\lambda_0$ and $\lambda_2$, with
$\lambda_1=\lambda_4=0$. The physically allowed region is determined
by the normalization condition
$\lambda_0^2+\lambda_2^2+\lambda_3^2=1$. The canonical W state,
corresponding to
$\lambda_0=\lambda_2=\lambda_3=1/\sqrt{3}$, is indicated by the
red star and gives the maximum value
$E=2\sqrt{2}/3\approx0.9428$.}
\label{fig:w_state}
\end{figure}

%\begin{figure}[htbp]
%\centering
%\includegraphics[width=\columnwidth]{Type_2b__E,_l0_.pdf}
%\caption{Variation of the measure ${E_{\mathrm{GHZ}}}
%= 2\lambda_0\lambda_4$ as a function of $\lambda_0$ for Type 2(b) GHZ-type states in the Acín classification. The maximum occurs at the maximally entangled GHZ state.}
%\label{fig:ghz_class}
%\end{figure}
{As shown in Fig.~\ref{fig:w_state}, the measure remains nonzero throughout
the interior of the W-type representative slice. The maximum on this symmetric slice
occurs at the canonical W representative,}
\begin{equation}
\lambda_0=\lambda_2=\lambda_3=\frac{1}{\sqrt{3}},
\end{equation}
{for which}
\begin{equation}
E_W=\frac{2\sqrt{2}}{3}\approx 0.943.
\label{eq:E_W_canonical_value}
\end{equation}
{The nonzero value of $E$ for W-type states reflects the fact that $E$ and the three-tangle quantify
different aspects of multipartite correlations. The three-tangle detects GHZ-type
genuine tripartite entanglement, while $E$ measures the average one-qubit-versus-rest
entanglement.}

{To quantitatively examine the behavior of the measures across different
entanglement classes, we consider a continuous one-parameter path in
the Acín parameter space. The path connects representative product,
biseparable, W, and GHZ states. The corresponding Acín coefficient
vectors
$\boldsymbol{\lambda}
=(\lambda_0,\lambda_1,\lambda_2,\lambda_3,\lambda_4)$ are chosen as
\begin{equation}
\begin{aligned}
\boldsymbol{\lambda}^{(P)}
&=(1,0,0,0,0),\\
\boldsymbol{\lambda}^{(B)}
&=\left(\tfrac{1}{\sqrt{2}},0,
\tfrac{1}{\sqrt{2}},0,0\right),\\
\boldsymbol{\lambda}^{(W)}
&=\left(\tfrac{1}{\sqrt{3}},0,
\tfrac{1}{\sqrt{3}},\tfrac{1}{\sqrt{3}},0\right),\\
\boldsymbol{\lambda}^{(GHZ)}
&=\left(\tfrac{1}{\sqrt{2}},0,0,0,
\tfrac{1}{\sqrt{2}}\right).
\end{aligned}
\label{eq:representative_states}
\end{equation}
A normalized linear interpolation between successive representative states is then defined as
\begin{equation}
\boldsymbol{\lambda}(s)=
\frac{(1-t)\boldsymbol{\lambda}^{(i)}
+t\boldsymbol{\lambda}^{(i+1)}}
{\left\|(1-t)\boldsymbol{\lambda}^{(i)}
+t\boldsymbol{\lambda}^{(i+1)}\right\|},
\label{eq:interpolation_path}
\end{equation}
where $t\in[0,1]$ is the local interpolation parameter and is related
to the global path parameter $s\in[0,1]$ by
\begin{equation}
t=
\begin{cases}
3s, & 0\leq s\leq\frac{1}{3},\\
3s-1, & \frac{1}{3}<s\leq\frac{2}{3},\\
3s-2, & \frac{2}{3}<s\leq1.
\end{cases}
\label{eq:interpolation_t}
\end{equation}
Thus, the three segments correspond to
$P\rightarrow B$, $B\rightarrow W$, and $W\rightarrow GHZ$,
respectively. Throughout the interpolation, we take
$\lambda_1=0$ and fix the phase parameter to $\phi=0$.
The normalization of the interpolated coefficient vector ensures
$\sum_{i=0}^{4}\lambda_i^2=1$ at every point along the path.
The resulting coefficients $\lambda_i(s)$ are then substituted into
the expressions for $E$, $\tau_3$, and $\mathcal{N}$ to evaluate
their behavior along the path, as shown in Fig.~\ref{fig:comparison_all_class}.

{Interestingly, the three-tangle $\tau_3$ is non-zero only for the GHZ class states, whereas the tripartite-negativity $\mathcal{N}$ vanishe for the  biseparable states. 
\begin{figure}[htbp]
\centering
\includegraphics[width=\columnwidth]{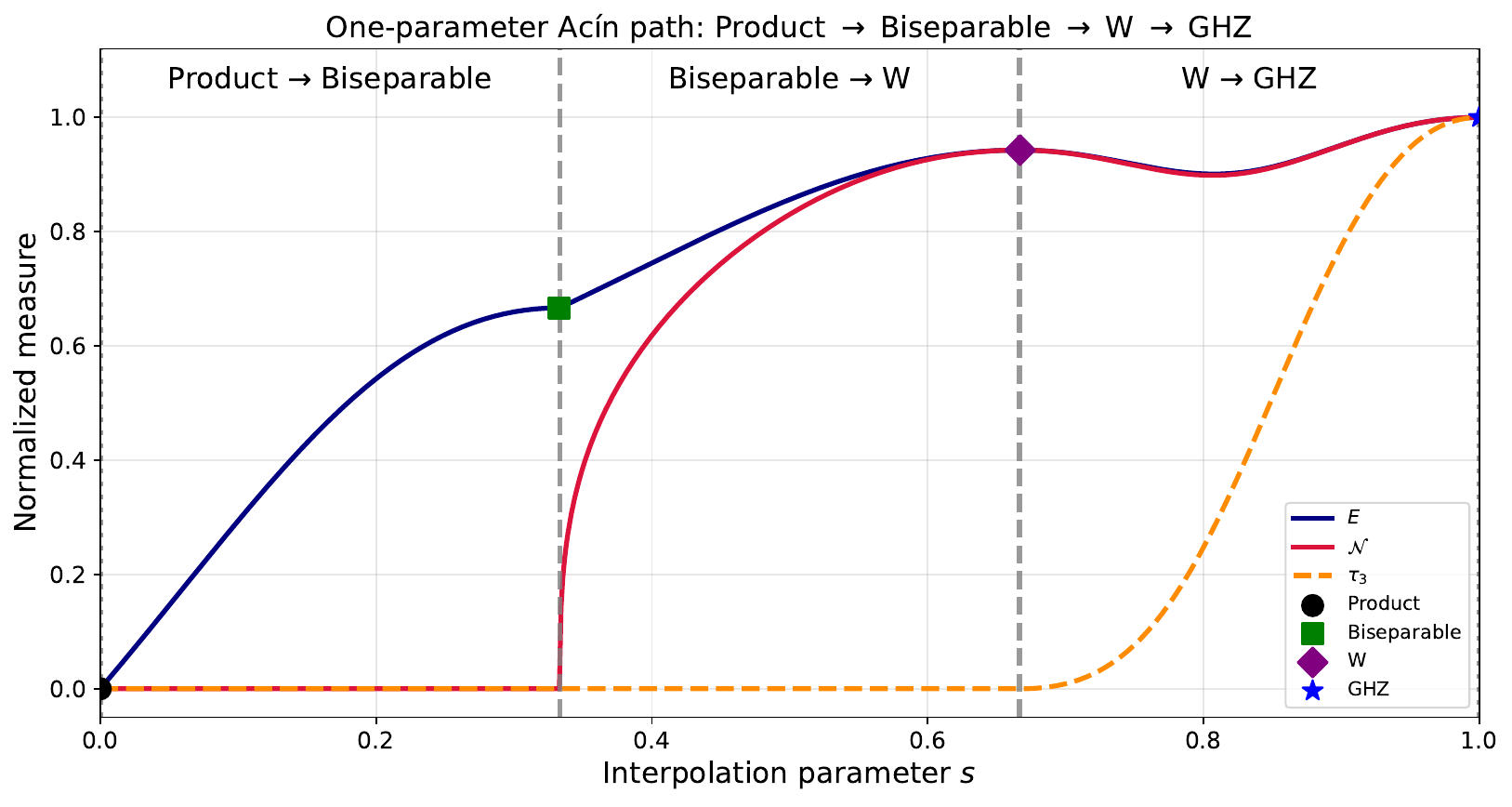}
\caption{Comparison of the global concurrence $E$, tripartite
negativity $\mathcal{N}$, and three-tangle $\tau_3$ along a
piecewise linear path $P\rightarrow B\rightarrow W\rightarrow GHZ$
in the Acín coefficient space. The path connects representative
product, biseparable, W-type, and GHZ-type states. The measures
exhibit distinct behavior along the path: $E$ remains nonzero in the
biseparable region, whereas $\mathcal{N}$ and $\tau_3$ vanish when
genuine tripartite entanglement is absent.}
\label{fig:comparison_all_class}
\end{figure}

\subsection{Discussion}

{The numerical results support the following observations.}

\begin{enumerate}
    \item \textbf{{Normalization:}} 
    {The plotted quantity satisfies $0\leq E\leq 1$. The maximally entangled GHZ
    state reaches $E=1$.}

    \item \textbf{{Product states:}} 
    {The measure vanishes for fully separable product states, as all one-qubit
    reduced density matrices are pure and hence have zero determinant.}

    \item \textbf{{Biseparable states:}} 
    {For biseparable states with an entangled two-qubit subsystem, $E$ can be
    nonzero. This confirms that $E$ detects bipartite
    entanglement embedded in a three-qubit state.

    \item \textbf{{W-type states:}} 
    {For the W-type representative family $\lambda_1=\lambda_4=0$, the canonical
    W point gives}
    \begin{equation}
    E_W=\frac{2\sqrt{2}}{3}.
    \end{equation}
    {The nonzero value of $E$ for W-type states reflects nonzero one-qubit-versus-rest
    entanglement.}

    \item \textbf{{GHZ-type states:}} 
    {For the GHZ representative}
    \begin{equation}
    |\psi_{\rm GHZ}\rangle=\lambda_0|000\rangle+\lambda_4|111\rangle,
    \end{equation}
    {the measure reduces to}
    \begin{equation}
    E_{\rm GHZ}=2\lambda_0\lambda_4,
    \end{equation}
    {and reaches $E=1$ for $\lambda_0=\lambda_4=1/\sqrt{2}$.}

    \item \textbf{Smooth variation:} The measure varies continuously across the Acín parameter space, making it suitable for optimization and analysis of mixed entanglement types.

\item \textbf{Phase independence:} For most classes in the Acín classification, the measure is independent of the phase parameter $\phi$, simplifying the analysis.

    \item \textbf{Complementarity with other measures:}
    The comparison with $\tau_3$ and $\mathcal{N}$ shows that these measures
    quantify different aspects of multipartite correlations. The three-tangle is
    sensitive to GHZ-type genuine tripartite entanglement, while $E$ quantifies the
    average one-qubit-versus-rest entanglement.}
\end{enumerate}

\subsection{Comparison with three-tangle}

{We now compare the measure $E$ with the three-tangle $\tau_3$.
The three-tangle, introduced by Coffman, Kundu, and Wootters~\cite{Coffman2000},
is a measure of GHZ-type genuine tripartite entanglement for three-qubit pure states.
For the Acín canonical form, it is given by}
\begin{equation}
\tau_3=4\lambda_0^2\lambda_4^2.
\label{eq:three_tangle_acin}
\end{equation}

{The comparison is useful because $E$ and $\tau_3$ quantify different aspects
of multipartite correlations. The quantity $E$ is the arithmetic mean of the three
one-qubit-versus-rest concurrences,}
\begin{equation}
E=\frac{1}{3}
\left(
C_{A|BC}+C_{B|AC}+C_{C|AB}
\right),
\label{eq:E_average_concurrence_again}
\end{equation}
{whereas $\tau_3$ detects the residual GHZ-type component that cannot be
accounted for by pairwise entanglement. Therefore, a nonzero value of $E$ when
$\tau_3=0$ should not be interpreted as superiority of $E$ over $\tau_3$. Rather,
it reflects that the two quantities measure different physical features.}

\subsubsection{GHZ-class state}

To compare the behavior of E with the standard three-tangle, we first consider the Type 2(b) GHZ-class state. For this case, both measures depend on the same state parameter $\lambda_0$ (with $\lambda_4 = \sqrt{1-\lambda_0^2}$). 
It is observed that both measures become zero at the separable points ($\lambda_0 = 0$ or $\lambda_0 = 1$) and reach their maximum value at the maximally entangled GHZ state with $\lambda_0 = 1/\sqrt{2}$.
However, if we compare the behavior independent of the scale, it can be seen that the variation of E is slightly different. In particular, in the lower parameter region, the value of the three-tangle increases more slowly, while E increases more rapidly. 
\subsubsection{W-class state}

For the W-class state (Type 3a), a clear difference can be observed. The value of the three-tangle remains zero for all parameter values in this class.
On the other hand, E remains non-zero for the same state. 

This indicates that global concurrence is sensitive to bipartite entanglement contributions and provides additional information compared to the three-tangle.

\subsubsection{Summary of comparison}

{The comparison between $E$ and the three-tangle may be summarized as follows:}

\begin{itemize}
    \item \textbf{{GHZ-type representatives:}}
    {Both $E$ and $\tau_3$ detect the entanglement of the generalized GHZ
    representative. They vanish at the product endpoints and reach their maximum values
    at the maximally entangled GHZ state. However, their functional dependence differs:
    $E_{\rm GHZ}=2\lambda_0\lambda_4$, while
    $\tau_3=4\lambda_0^2\lambda_4^2$.}

    \item \textbf{{W-type representatives:}}
    {For W-type states, $\tau_3=0$, while $E$ remains nonzero. This reflects
    the fact that the W state has nonzero one-qubit-versus-rest entanglement but zero
    GHZ-type three-tangle.}

    \item \textbf{{Complementarity:}}
    {The three-tangle identifies GHZ-type genuine tripartite entanglement,
    whereas $E$ quantifies the average marginal entanglement across the three
    one-qubit-versus-rest cuts. Therefore, the two measures provide complementary
    information about the structure of three-qubit correlations.}

    \item \textbf{{Limitation of $E$:}}
    {Because $E$ is nonzero for certain biseparable states, it should not be
    interpreted as a genuine tripartite entanglement measure alone. Instead, it detects
    total one-qubit-versus-rest entanglement, including embedded bipartite contributions.}
\end{itemize}

{This comparison directly addresses the distinction between genuine tripartite
entanglement and bipartition-based entanglement. The nonzero value of $E$ for W-type
or biseparable states is therefore a feature of its definition.}

\subsection{Comparison with tripartite negativity}

To further examine the effectiveness of global concurrence and evaluate its relation to other established multipartite entanglement quantifiers, we now compare its behavior with tripartite negativity. 
Tripartite negativity, introduced as an extension of bipartite negativity to multipartite systems \cite{Sabin2008}, is based on the geometric mean of negativities across all bipartitions of a three-qubit state. Unlike the three-tangle, tripartite negativity can detect entanglement in both GHZ-class and W-class states, making it a broader entanglement detector.

Mathematically, tripartite negativity is defined as
\begin{equation}
    \mathcal{N} = (N_{A-BC}N_{B-AC}N_{C-AB})^{1/3},
\end{equation}
where each term corresponds to the bipartite negativity for a particular partition.
Since both tripartite negativity $\mathcal{N}$ and global concurrence $E$ are constructed from bipartition-related properties of the state, comparing them provides important insight into whether global concurrence captures similar or additional entanglement features.

\subsubsection{GHZ-class state}

{For the GHZ-class states (Type 2(b)), both $E$ and $\mathcal{N}$ depend on the same parameter $\lambda_0$, with $\lambda_4 = \sqrt{1-\lambda_0^2}$.
Fig.~\ref{fig:comparison_tripartite_negativity}(a) illustrates the variation of $E_{\mathrm{GHZ}}
= 2\lambda_0\lambda_4$ and $\mathcal{N}$ for the Type 2(b) GHZ class as a function of $\lambda_0$. Both measures vanish at $\lambda_0 = 0$ and $\lambda_0 = 1$ (product states) and reach their maximum values of 1 at $\lambda_0 = 1/\sqrt{2}$, corresponding to the maximally entangled GHZ state.}

{Thus, both measures display identical  dependence on $\lambda_0$ for GHZ-class states, consistently detecting strong multipartite entanglement.}
\begin{figure}[h]
\centering
\includegraphics[width=\columnwidth]{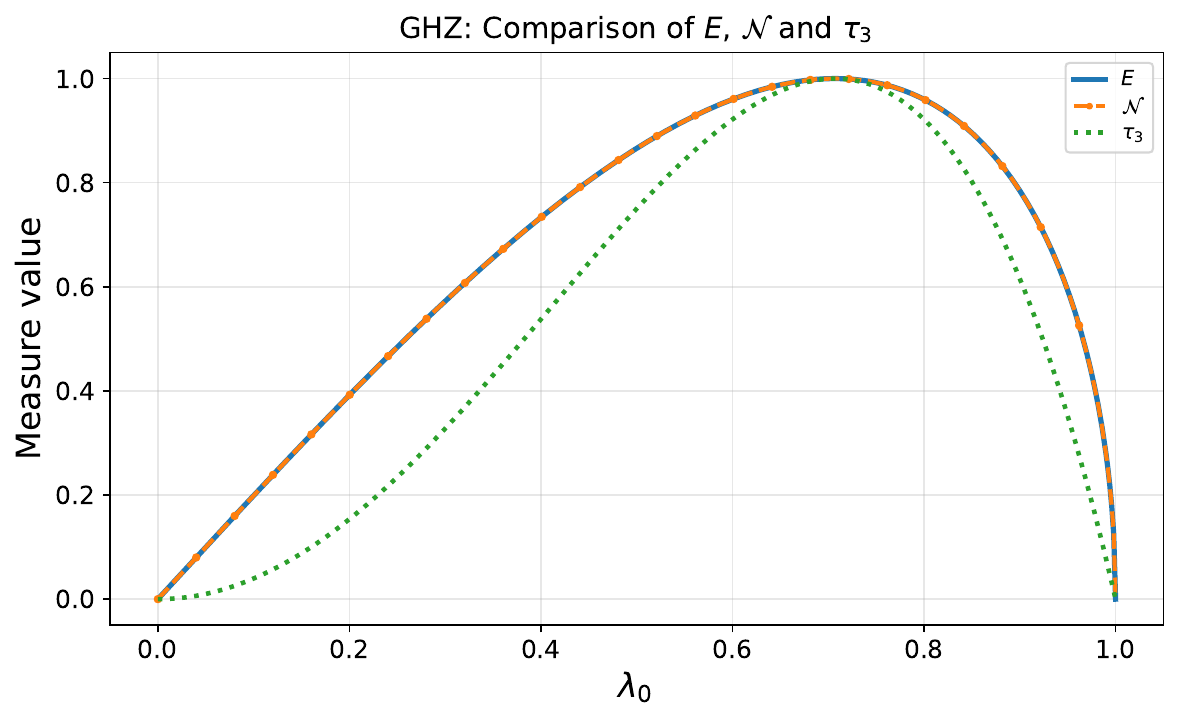}
\\
\includegraphics[width=\columnwidth]{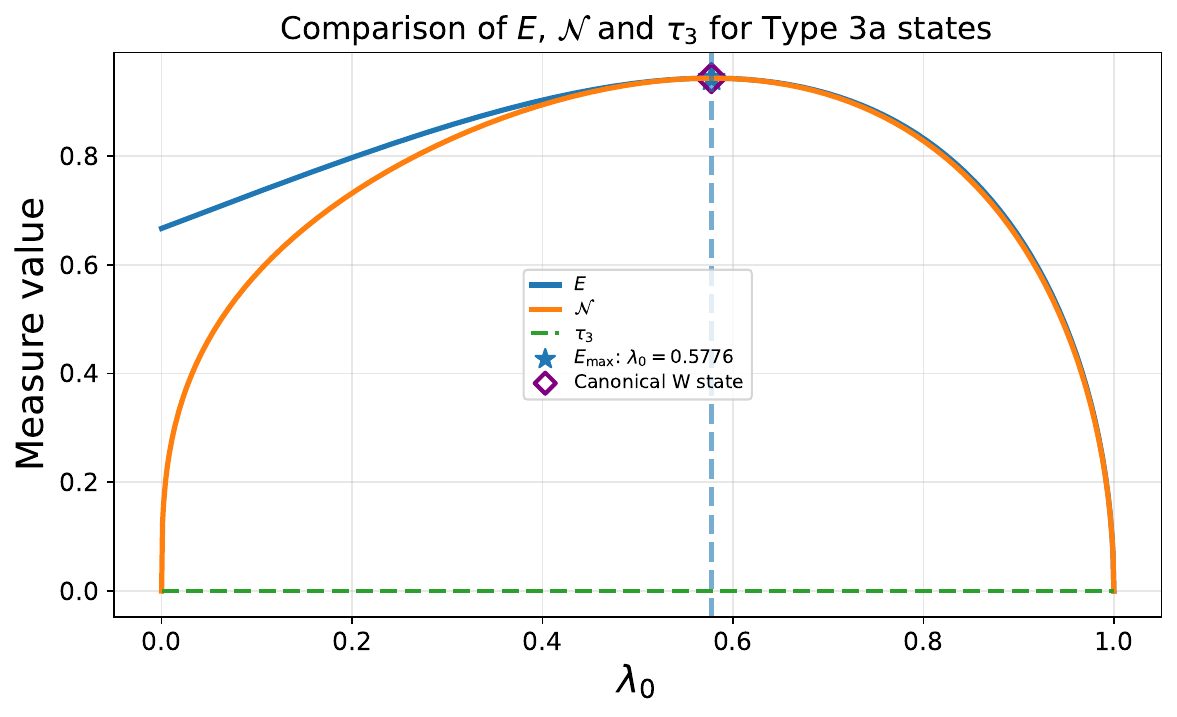}
\caption{{Comparison of the global concurrence $E$ with tripartite negativity $\mathcal{N}$ and three-tangle $\tau_3$ for (a) GHZ-class state and (b) W-class state. For GHZ states, $E$ and $\mathcal{N}$ show virtually identical behavior. $\tau_3$ differs substantially from the other two. For W states, comparison reveals that global concurrence captures broader bipartite entanglement redistribution, while tripartite negativity is maximized at the canonical W state. Three-tangle remains zero throughout the parameter range.}}
\label{fig:comparison_tripartite_negativity}
\end{figure}

\subsubsection{W-class state}

For the W-type states (Type 3(a)), the comparison between the global
concurrence and tripartite negativity reveals a distinct behavior of
the two measures. Although both measures attain their maximum at the
canonical W state, their magnitudes and behavior away from this point
differ, reflecting the different aspects of entanglement captured by
the two measures.

The tripartite negativity $\mathcal{N}$ reaches its maximum at the
canonical W state and decreases as the state approaches the
biseparable boundary. It vanishes at $\lambda_0=0$, where genuine
tripartite entanglement is absent. In contrast, the global concurrence
remains nonzero at this boundary because bipartite entanglement can
still persist within the two-qubit subsystem. Thus, the global
concurrence remains sensitive to entanglement even when the genuine
tripartite component vanishes.

The comparison therefore highlights a complementary behavior of the
two measures: tripartite negativity is sensitive specifically to
genuine tripartite entanglement, whereas the global concurrence also
captures bipartite entanglement present in biseparable states.

\section{Conclusion}

{In this work, we studied the Bhaskara--Panigrahi global concurrence for
pure three-qubit states in the Acín canonical parametrization. We derived 
expressions for the one-qubit reduced density matrices and their determinants, leading
to an explicit formula for the normalized measure}
\begin{equation}
E
=
\frac{2}{3}
\left(
\sqrt{\det\rho_A}
+
\sqrt{\det\rho_B}
+
\sqrt{\det\rho_C}
\right).
\end{equation}
{Equivalently, $E$ can be written as}
\begin{equation}
E
=
\frac{1}{3}
\left(
C_{A|BC}+C_{B|AC}+C_{C|AB}
\right),
\end{equation}
{where $C_{A|BC}$, $C_{B|AC}$, and $C_{C|AB}$ are the bipartite concurrences
associated with the three one-qubit-versus-rest partitions. This representation gives
a clear physical interpretation of $E$ as the average one-qubit-versus-rest entanglement.}

{We established several basic properties of the measure. In particular, $E$ is
nonnegative, normalized according to $0\leq E\leq 1$, invariant under local unitary
transformations, symmetric under permutations of the parties, and non-increasing on
average under LOCC for pure three-qubit states. The maximum value $E=1$ is reached
by the maximally entangled GHZ state.}

{At the same time, the analysis clarifies important limitations. The
measure is not a SLOCC invariant and therefore cannot serve as a complete classifier
of SLOCC equivalence classes. Moreover, because $E$ can be nonzero for biseparable
states with an entangled two-qubit subsystem, it should not be interpreted as a measure
of genuine tripartite entanglement alone. Instead, it quantifies the total average
entanglement between each single qubit and the remaining pair.}

{Representative Acín parameter families were analyzed to illustrate the behavior
of the formula. The measure vanishes for product states, is nonzero for
biseparable states containing bipartite entanglement, gives $E=1$ for the maximally
entangled GHZ state, and gives $E=2\sqrt{2}/3$ for the canonical W representative.
Comparisons with the three-tangle and tripartite negativity show that these quantities
capture complementary aspects of three-qubit correlations. The three-tangle detects
GHZ-type genuine tripartite entanglement, whereas $E$ captures average bipartition
entanglement across the three one-qubit cuts.}

{The present analysis is restricted to pure three-qubit states.
Nevertheless, the pure-state expression obtained here provides a natural
starting point for extending the measure to mixed states through the
convex-roof construction. For a mixed state $\rho$, one may define
\begin{equation}
E(\rho)
=
\min_{\{p_i,|\psi_i\rangle\}}
\sum_i p_i E(|\psi_i\rangle),
\qquad
\rho=\sum_i p_i|\psi_i\rangle\langle\psi_i|,
\end{equation}
where the minimization is taken over all pure-state decompositions of
$\rho$. The evaluation of this minimization and the resulting properties
of the mixed-state extension are beyond the scope of the present work and
may be investigated in future studies.}

From an application viewpoint, the expression for \(E\) provides a simple nonlinear diagnostic for tracking the distribution of entanglement across the three one-qubit bipartitions of a pure three-qubit state. This may be useful in numerical simulations of entanglement generation, quantum-state preparation, decoherence-induced redistribution of correlations, and comparative studies of GHZ-type and W-type entanglement resources.

Overall, the work contributes to the ongoing effort to better understand and quantify multipartite entanglement in quantum information theory. The analytical simplicity of the measure, suggests that it may serve as a useful tool for future studies of entanglement structure, and applications in quantum information processing. Future investigations may extend this approach to mixed states, higher-dimensional multipartite systems, and dynamical scenarios involving decoherence and entanglement evolution.

\begin{acknowledgments}
The author sincerely thanks Prof. Tabish Qureshi for his guidance, supervision, and the Centre for Theoretical Physics, Jamia Millia Islamia, for providing the facilities and academic environment in which this work was completed.
\end{acknowledgments}

\section*{Declarations}
\noindent \textbf{Data availability:} No additional data was generated in this work.\\
\textbf{Funding} No funding was used in this work.\\
\textbf{Competing interests:} The author declares there are no competing interests.\\
\textbf{Authors Contributions:} Not applicable.

% ==================================================
% Bibliography
% ==================================================
\FloatBarrier

\end{document}